\documentclass[nofootinbib,prb,amsmath,amssymb,twocolumn,floatfix]{revtex4-2}

\usepackage{graphicx}
\usepackage{dcolumn}
\usepackage{bm}
\usepackage{siunitx}
\usepackage{mathtools}
\usepackage{xcolor}
\usepackage{yhmath}
\usepackage[hidelinks]{hyperref}
\usepackage{todonotes}
\graphicspath{ {images/} }

\newcommand{\B}{\mathcal{B}}

\newcommand{\R}{\mathbb{R}}

\newcommand{\Z}{\mathbb{Z}}

\newcommand{\p}{\partial}
\newcommand{\dx}{\: \mathrm{d}}

\renewcommand{\b}[1]{\textbf{#1}}

\newcommand{\ds}{\displaystyle}
\newcommand{\iu}{\mathrm{i}\mkern1mu}

\newcommand{\sddots}{\scalebox{.5}{$\ddots$}}
\newcommand{\sadots}{\scalebox{.5}{$\adots$}}

\begin{document}
	
	\preprint{APS/123-QED}
	
	\title{Perturbation theory for dispersion relations of spacetime-periodic materials}
	
	\author{E.~O. Hiltunen$^{1}$}
	\affiliation{$^1$Department of Mathematics, University of Oslo, Moltke Moes vei 35, 0851 Oslo, Norway}
	
	\date{\today}
	
	\begin{abstract}
		We consider Bloch states of weak spacetime-periodic perturbations of homogeneous materials in one spatial dimension. The interplay of space- and time-periodicity leads to an infinitely degenerate dispersion relation in the free case. We consider a general perturbation term, and, as consequence of the infinite degeneracy, we show that the effective equations are given by the eigenvalue problem of an infinite matrix. Our method can be viewed as a time-modulated generalisation of the nearly-free electron model. Based on this result, we find that the infinite degeneracy may split into a family of non-degenerate bands. Our results are illustrated with numerical calculations, and we observe close agreement between the perturbation theory and the numerically computed full solution.
	\end{abstract}
	
	\maketitle
	
	\section{Introduction}
	The field of time-dependent materials has gained a wealth of recent interest due to promising applications and experimental progress \cite{galiffi2022photonics,koutserimpas2023multiharmonic,tirole2023double, peng2016experimental,caloz2019spacetime1,caloz2019spacetime2,mendoncca2002time, koutserimpas2018electromagnetic,yu2022band,sotoodehfar2022waves,rizza2022short}. Many phenomena of metamaterials, where exotic wave effects are realised through a small-scale periodic spatial structure, can be transposed to time-dependent structures. Notably, such structures might exhibit momentum gaps, in direct analogy to frequency band gaps in conventional metamaterials. Coupling time-modulation to conventional metamaterials (so called \emph{spacetime} metamaterials or \emph{Floquet} metamaterials) offers an additional degree of freedom, allowing exciting effects such as non-reciprocal transmission \cite{simon1960action, oliner1961wave, ourir2019active}, Fresnel drag \cite{huidobro2019fresnel}, frequency conversion \cite{yin2022floquet}, parametric amplification \cite{cullen1958travelling, raiford1974degenerate}, exceptional points \cite{nikzamir2022achieve,rouhi2020exceptional}, and topological localisation \cite{rechtsman2013photonic,fleury2016floquet}. In this way, spacetime-periodic structures are pushing the boundaries of wave control and represents the current frontier in metamaterial design.
	
	A well-studied class of spacetime metamaterials are synthetically moving gratings, where the material parameters are piecewise constant and uniformly translated in time \cite{pendry2021gain,huidobro2019fresnel, nassar2017modulated, huidobro2021homogenization,touboul2024high,lurie2007introduction}. Such materials offer a natural platform for non-reciprocal wave transmission, due to their inherent breaking of time-reversal symmetry. Considering more general time-varying materials, a simplifying assumption is the adiabatic limit of slow modulation, under which perturbation theories can be developed \cite{weinberg2017adiabatic,arkhipov2024restoring,menyuk1985particle,young1970adiabatic}. Nevertheless, key features of spacetime materials arise from the coupling between the space- and time-varying components.	While the band structure of other spacetime-periodic materials have been studied in a number of works (see, for example, \cite{sagiv2022effective,hameedi2023radiative,ammari2021time, feppon2024subwavelength, ammari2023transmission,ammari2024scattering,asadchy2022parametric}), a general mathematical analysis of spacetime-periodic materials is currently beyond reach.
		
	The main features of spacetime metamaterials are governed by the interplay between the space- and time periodicities. As an example of this principle, field pattern materials is a class of materials arising when these periodicities are commensurable with respect to the wave speed of the constituent phases  \cite{mattei2023effects,mattei2018field,mattei2017field}. In sharp contrast to conventional metamaterials, field pattern materials have been shown to have infinitely-degenerate band structure. The time-periodicity induces a coupling between Fourier harmonics of different frequencies, which folds the frequency axis and, under suitable conditions of the spacetime-periodicity, causes bands to coalesce. Such high degeneracies do not appear in moving-grating materials, where the Fourier harmonics decouple after transforming into a moving coordinate frame. 
	
\begin{figure*}[tbh]
		\centering
		\includegraphics[width=\linewidth]{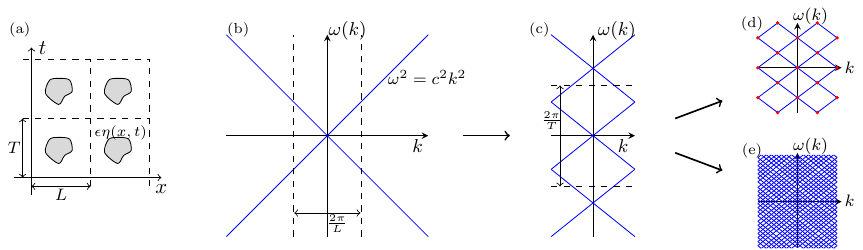}
		\caption{We consider a $(L,T)$-spacetime-periodic material whose refractive index $n$ satisfies $n^2(x,t) = 1+\epsilon\eta(x,t)$ for $\epsilon\ll 1$, sketched in (a). The dispersion of the free problem, shown in (b), is given by $\omega^2 = c^2k^2$. For spacetime-periodic media in the perturbation regime, the free dispersion is folded in both the momentum (c) and frequency (d),(e) variables. If $\alpha = L/cT$ is rational (shown in the case $\alpha = 3/2$ in (d)), the folded bands coalesce, and we develop an infinitely degenerate perturbation theory to compute the perturbed bands. The band intersections, marked by red points in (d), are higher-order degeneracies around which a different perturbation theory holds. If, on the other hand, $\alpha$ is irrational, the folded bands fill the entire $(k,\omega)$-space (e). Although each degeneracy has multiplicity 2 in the latter case, each band is arbitrarily close to other bands. Because of this, we describe the irrational case by a sequence of rational approximants. }\label{fig:folding}
	\end{figure*}

	In this work, we develop a perturbation theory for infinitely-degenerate band structures of spacetime-periodic materials. The method is elementary in nature, yet has profound impact on our understanding of Floquet metamaterials. We consider a scalar wave model with a weak perturbation term $\epsilon\eta$ (for small $\epsilon$) of a homogeneous, static material. The perturbation $\eta$, sketched in FIG. \ref{fig:folding}(a), is periodic in both space and time. We emphasise that no other assumptions are imposed on $\eta$. Guided by recent experimental progress in the setting of photonic and phononic systems, we consider a time-modulated scalar wave equation which readily models photonic or acoustic systems. In the language of Schrödinger's equation, this perturbation theory is a time-modulated extension of the well-known nearly-free electron model. The infinitely-degenerate eigenstates of the free problem gives rise to an effective system of equations which, albeit discrete, is infinite-dimensional. The coupling terms are given by the Fourier coefficients of $\eta$, and, if $\eta$ is smooth in $t$, the system can be truncated with good accuracy. Moreover, under an additional regularity assumption, we are able to explicitly solve the effective equations.
	
	We will develop the method for a scalar wave model in $(1+1)$-dimensions:
	\begin{equation}\label{eq:wave}
		\left(\frac{\p^2}{\p t^2} - \frac{c^2}{n^2(x,t)}\frac{\p^2}{\p x^2} \right) u(x,t) = 0, \quad x,t\in \R.
	\end{equation}
	with wave speed $c$ and refractive index $n(x,t)$. The goal of the work is to study the perturbation theory when $n^2$ is a shallow perturbation of a constant background:
	\begin{equation}
		n^2(x,t) = 1+\epsilon \eta(x,t)
	\end{equation}
	for $\epsilon \ll 1$. We assume $\eta(x,t)$ to be $L$-periodic in $x$ and $T$-periodic in $t$. The frequency $\Omega$ of time modulation is $\Omega = 2\pi/T$. Here, the quasifrequency $\omega$ is defined in the temporal Brillouin zone $\B_t = \R/\Omega\Z$ while the quasimomentum $k$ is defined in the spatial Brillouin zone $\B_x = \R/\tfrac{2\pi}{L}\Z$. The key parameter $\alpha$ is defined as
	\begin{equation}
		\alpha = \frac{L}{cT}.
	\end{equation}
As illustrated in FIG. \ref{fig:folding} (d), rational $\alpha$ implies that the free band structure (corresponding to $\epsilon = 0$) is infinitely degenerate, while irrational $\alpha$ means that each point is two-fold (finitely) degenerate and the folded bands fill the entirety of the $(k,\omega)$-space.
	
	\section{Infinitely-degenerate perturbation theory}\label{sec:pert}
	We now develop the perturbation theory in the case when $\alpha$ is rational:
	\begin{equation}
		\alpha = \frac{p}{q}, \quad p,q\in \Z, \quad \text{gcd}(p,q) = 1.
	\end{equation}
	The free case $\epsilon = 0$ is infinitely degenerate, and consequently, the perturbed band structure will be given by an eigenvalue problem in terms of an infinite matrix. As we shall see, the numbers $p$ and $q$, respectively, dictate the order of coupling between Fourier modes in the momentum and frequency variables. We begin by considering perturbations around points away from band intersections (marked as red points in FIG. \ref{fig:folding}(d)) and present the higher-degenerate perturbation theory in Section \ref{sec:higher}.
	
	By Bloch's theorem in the temporal variable, we seek solutions to \eqref{eq:wave} which are $\omega$-quasiperiodic in time $t$:
	\begin{equation}\label{eq:un}
		u(x,t) = \sum_{n=-\infty}^\infty u_n(x) e^{-\iu (\omega+n\Omega)t},
	\end{equation}
	for some quasifrequency $\omega \in \B_t$.	Substituting into \eqref{eq:wave} leads to a coupled system of equations for the coefficients $u_n(x)$, given by
	\begin{equation}\label{eq:spatial}
		\left(\partial_{xx} + \tfrac{(\omega+n\Omega)^2}{c^2}\right)u_n + \epsilon\sum_{m=-\infty}^\infty\hat{\eta}(x,n-m)\tfrac{(\omega+m\Omega)^2}{c^2} u_m =0.
	\end{equation}
	where $\hat{\eta}(x,n)$ are the one-dimensional Fourier coefficients of $\eta(x,t)$ in $t$ given by
	\begin{equation}
		\hat{\eta}(x,n) = \frac{1}{T}\int_0^Te^{\iu n \frac{2\pi}{T}t}\eta(x,t) \dx t.
	\end{equation}
	We adopt a (formal) asymptotic series solution of $u_n$ and $\omega$ as 
	\begin{align}
		u_n = u_n^{(0)} + \epsilon u_n^{(1)} + ... \label{eq:seriesu}\\
		\omega = \omega^{(0)} + \epsilon \omega^{(1)} + ...	 \label{eq:serieso}
	\end{align}
	Here, $\bigl(k,\omega^{(0)}(k)\bigr)$ is a point on the (folded) free dispersion diagram (c.f. FIG. \ref{fig:folding}(d)) while $u^{(0)}$ is corresponding eigenmode. We substitute \eqref{eq:seriesu}--\eqref{eq:serieso} into \eqref{eq:spatial} and separate by order of $\epsilon$. At order $O(1)$, we have 
	\begin{equation}\label{eq:e0}
		\partial_{xx} u_n^{(0)} + \frac{(\omega^{(0)}+n\Omega)^2}{c^2}u_n^{(0)} = 0.
	\end{equation}
 	Crucially, in order to satisfy the (spatial) Bloch condition for some quasimomentum $k\in \B_x$, the integer $n$ must be divisible by $q$. For $j\in \Z$, we then have
	\begin{equation}\label{eq:u0}
		u_n^{(0)}(x) = \begin{cases}a_j v_j(x), \quad &n = qj, \\ 0, & \text{otherwise}, \end{cases}
	\end{equation}
	where $v_j := e^{\pm \iu \frac{\omega^{(0)}+qj\Omega}{c}x},$ and $\omega^{(0)}(k) = c(\pm k+\frac{2\pi \ell}{L})$ for some $\ell \in \Z$. Here, the sign depends on the perturbed point of the free dispersion. The coefficients $a_j$ will be determined by passing to higher orders in $\epsilon$.
	
	At order $O(\epsilon)$, we have 
	\begin{multline}\label{eq:Oe}
		\partial_{xx} u_n^{(1)} + \frac{(\omega^{(0)}+n\Omega)^2}{c^2}u_n^{(1)} + \frac{2\omega^{(1)}(\omega^{(0)} + n\Omega) }{c^2}u_n^{(0)} \\
		+ \sum_{m =-\infty}^\infty \hat{\eta}(x,n-m) \frac{(\omega^{(0)}+m\Omega)^2}{c^2} u_m^{(0)}  = 0.
	\end{multline}
	For $m,n\in \Z$, we adopt the following convention of the two-dimensional Fourier coefficients  $\widetilde{\eta}(m,n)$ of $\eta(x,t)$:
	\begin{equation}
		\widetilde{\eta}(m,n) = \frac{1}{TL}\int_0^L\int_0^Te^{-\iu m \frac{2\pi}{L}x}e^{\iu n \frac{2\pi}{T}t}\eta(x,t) \dx t \dx x.
	\end{equation}
	We define the matrix entries
	\begin{equation}
		H_{j} = \widetilde{\eta}\bigl(\pm pj,qj\bigr), \quad O_j = \omega^{(0)} + qj\Omega,
	\end{equation}
	and define the (doubly infinite) Toeplitz matrix $\mathbf{H}$ and the diagonal matrix $\mathbf{O}$ as 
	\begin{equation}
	\mathbf{H} = \begin{psmallmatrix} \sddots& \sddots & \sddots  &  & \hspace{-5pt}\sadots   \\
		\sddots& H_0 & H_{-1} & H_{-2}&  \\
		\sddots & H_{1} & H_0 & H_{-1}  & \sddots \\
		 & H_{2} & H_{1} & H_0  & \sddots \\
		\sadots &  & \sddots  & \sddots & \sddots \end{psmallmatrix}, \
	\mathbf{O} = \begin{psmallmatrix} \sddots \hspace{-3pt}&  &   &  &    \\
		 & O_{-1} \hspace{-5pt} & &  &   \\
		 & & O_{0}\hspace{-2pt} & &  \\
		 & & & O_1 &  \\[-0.3em]
		 &  &   &  & \hspace{-3pt}\sddots \end{psmallmatrix}.
	\end{equation}	
	Projecting \eqref{eq:Oe} onto the free eigenbasis $v_j(x)$, we obtain the matrix equation	
	\begin{equation}\label{eq:eval}
		-\frac{1}{2}\mathbf{H}\mathbf{O}\mathbf{v} = \omega^{(1)}\mathbf{v},
	\end{equation}
	where $\mathbf{v} = \mathbf{O}\begin{psmallmatrix} \scalebox{.5}{$\vdots$} \\[-0.1em] a_{-1}\\a_0\\a_1\\[-0.2em] \scalebox{.5}{$\vdots$} \end{psmallmatrix}$. We are able to solve the eigenvalue problem \eqref{eq:eval} by passing to the time-domain: define $v(t)$ and $h(t)$ as
	\begin{equation}\label{eq:vh}
	v(t) = \sum_{j=-\infty}^\infty v_j e^{-\iu (\omega^{(0)} + jq\Omega)t }, \quad	h(t) = \sum_{j=-\infty}^\infty H_je^{-\iu jq\Omega t },
	\end{equation} 
	then \eqref{eq:eval} is equivalent to the first-order boundary-value problem
	\begin{equation}\label{eq:evalODE}
	\begin{cases} \ds -\frac{\iu}{2}h(t)v'(t) = \omega^{(1)}v(t),\\
	v\left(t+\tfrac{T}{q}\right) = e^{\iu \omega^{(0)}\frac{T}{q} }v(t).\end{cases}
	\end{equation}	
	We emphasise that the Bloch condition of \eqref{eq:evalODE} is phrased  with reduced time-periodicity $T/q$. To proceed, we assume that $\mathbf{H}$ is diagonally dominant:
	\begin{equation}\label{eq:assump}
		|H_0| > \sum_{n\neq 0} |H_n|,
	\end{equation} 
	so that $h(t)$ has no zeros and the ODE \eqref{eq:evalODE} is regular. In this case, the general solution is given by 
	\begin{equation}
	v(t) = Ce^{2\iu\omega^{(1)}\int_0^t\frac{\dx \tau}{h}}.
	\end{equation}
	Finally, by imposing the Bloch condition of \eqref{eq:evalODE}, we obtain 
	\begin{equation}\label{eq:result}
	\omega^{(1)} = \frac{1}{2\frac{q}{T}\int_0^{T/q}\frac{\dx \tau}{h} }(\omega^{(0)} + qn\Omega),
	\end{equation}
	for any integer $n\in \Z$.	Equation \eqref{eq:result} is the main result of the perturbation theory, which explicitly gives us the perturbation terms $\omega^{(1)}$ of the band structure. We obtain an infinite family of distinct  (simple) perturbation terms $\omega^{(1)}$, indexed by $n \in \Z$. As result, the infinitely degenerate free band has split into an infinite family of simple bands.
	
	\section{Piecewise constant modulated composites}
	Here, we illustrate the results for materials which are piecewise constant in space but modulated in time.  {In this simplified setting, the band structure may be numerically computed by truncating the Fourier series in \eqref{eq:un} to length $K$, providing means to validate the 
	perturbation theory \eqref{eq:result}.} Specifically, we can follow \cite[Section V]{hiltunen2024coupled} and perform a change of coordinates which transforms \eqref{eq:spatial} into a system of coupled Helmholtz equations in the interior and exterior regions, respectively. This way, we validate \eqref{eq:eval} by numerically computing the band structure of \eqref{eq:spatial} without relying on the perturbation theory. Throughout, we perform the numerical calculations by truncating the Fourier series with $K=11$. 
	
\begin{figure}[tbh]
		\centering
		\includegraphics[width=\linewidth]{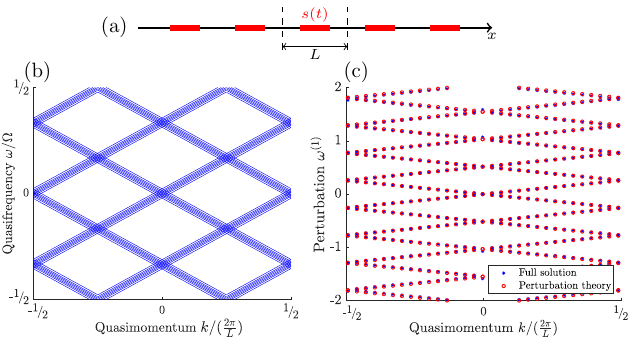}
		\caption{We perform numerical calculations for a spatially piecewise constant material. The geometry is sketched in (a), where we have a time-modulated inclusion in a static background. We consider the case $\alpha = 3/2$ and $\epsilon = 0.01$, where the full band structure is shown in (b). The perturbations $\omega^{(1)}$, computed either through the full solution of \eqref{eq:spatial} or through the perturbation theory \eqref{eq:result} are shown in (c), and we observe close agreement between the methods.}\label{fig:numerics1}
	\end{figure}

	To define the system, we take $\eta$ inside the central unit cell $x\in [-\tfrac{L}{2},\tfrac{L}{2}]$ as
	\begin{equation}\label{eq:eta}
		\eta(x,t) = \begin{cases}
			s(t), & |x| < \frac{a}{2},\\
			0, & \frac{a}{2} < |x| < \frac{L}{2},
	\end{cases}\end{equation}
	for some $T$-periodic function $s(t)$.  Equation \eqref{eq:eta} describes a time-modulated particle of width $a$ inside a static background medium, as sketched	 in FIG. \ref{fig:numerics1}(a). For simplicity, we take $s(t)$ to be a (finite) trigonometric polynomial. It is straight-forward to compute
	\begin{equation}
		\widetilde{\eta}(m,n) = \hat{s}(n)\begin{cases} \frac{1}{m\pi}\sin(\frac{\pi m a}{L}), \quad &m\neq 0, \\ \frac{a}{L}, & m=0. \end{cases}
	\end{equation} 
	
	In FIG. \ref{fig:numerics1}, we compute the band structure in the case $\alpha = 3/2$,  $\epsilon = 0.01$, and
	\begin{equation}
		s(t) = 1+\cos(\Omega t) + \cos(2 \Omega t).
	\end{equation}
	Observe that this choice of $\eta$ and $\alpha$ satisfies \eqref{eq:assump}. Comparing FIG. \ref{fig:numerics1}(b) with the free (folded) bands shown in FIG. \ref{fig:folding}(d), we see that the degenerate band of FIG. \ref{fig:folding}(d) open into a family of perturbed bands when $\epsilon \neq 0$. FIG. \ref{fig:numerics1}(c) shows the perturbation terms $\omega^{(1)}$, which are specified by the distance between the perturbed bands and the free bands. For the chosen value of $\epsilon$, we observe close agreement between the full solution and the perturbation theory.

\begin{figure}[tbh]
		\centering
		\includegraphics[width=0.8\linewidth]{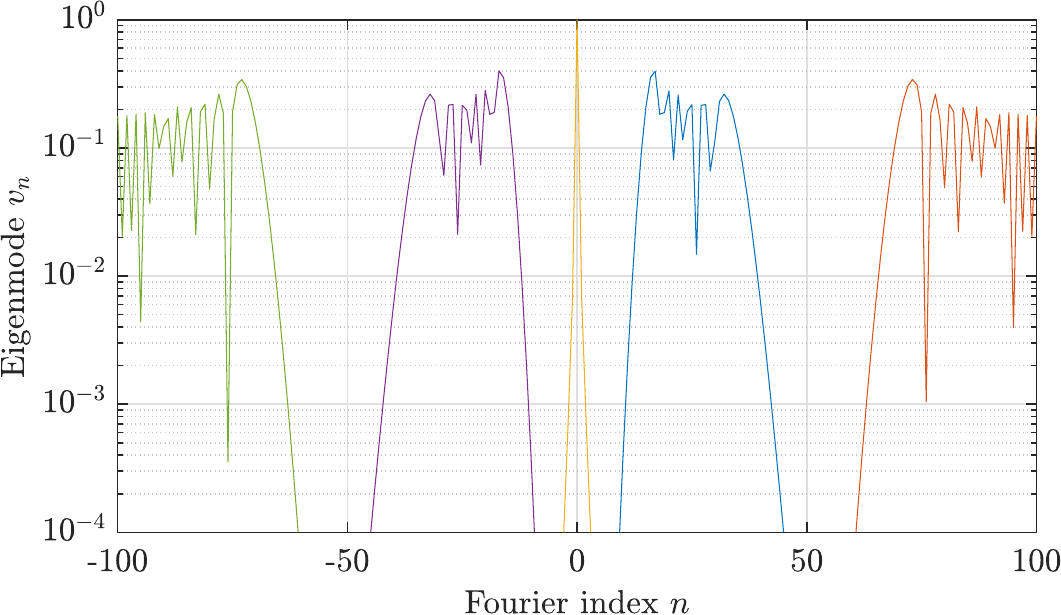}
		\caption{Under the assumption \eqref{eq:assump}, the eigenmodes are exponentially localized in the Fourier domain. Here, we plot a selection of the eigenmode solutions $v_n$ of \eqref{eq:eval} for large $K = 100$, and each mode decays exponentially to zero away from its support.}\label{fig:modes}
	\end{figure}

	It is informative to consider the eigenmodes of \eqref{eq:eval}. In FIG. \ref{fig:modes}, we plot a selection of the eigenmodes in the frequency domain. As expected, due to \eqref{eq:assump}, the solutions are smooth in $t$, hence supported on a finite set of Fourier modes and exponentially decaying away from this set.  

	\section{Travelling-wave modulation}
	 {Next, we illustrate how the perturbation theory can be applied to travelling-wave modulations of the form}
	\begin{equation}
		\eta(x,t) = f(gx-\Omega t),
	\end{equation}
	where $g=\frac{2\pi}{L}$. Here, $f$ is a $2\pi$-periodic function, describing the wave form of the modulation travelling by speed $v=\frac{\Omega}{g}$. Such modulation has received considerable interest in recent years, and the problem allows analytic solution by the change of coordinates
	\begin{equation}\label{eq:coord}
		\begin{cases}
			x' = x-vt,\\ t' = t,
		\end{cases}
	\end{equation}
	corresponding to a coordinate frame which moves along the travelling-wave modulation (see, for example, \cite{lurie2007introduction, galiffi2019broadband,pendry2021gain} for an extensive treatment of such problems). In particular, for frequency $\omega'$ and quasimomentum $k'$ in the moving coordinate frame, the corresponding quantities in the original coordinates are 
	\begin{equation}\label{eq:coord}
		\begin{cases}
			k = k',\\ \omega = \omega' + vk'.
		\end{cases}
	\end{equation}
	Note that $k'$ is defined modulo $g$. This means that $\omega$ is defined modulo $\Omega$, consistently with the folding shown in FIG.~\ref{fig:folding}. Following the perturbation theory of Section \ref{sec:pert}, we can compute 
	\begin{equation}
		\widetilde{\eta}(m,n) = \delta_{mn}f_n,
	\end{equation}
	where $\delta_{mn}$ is the Kroenecker delta and $f_n$ are the Fourier coefficients of $f$.
	
\begin{figure}[tbh]
		\centering
		\includegraphics[width=\linewidth]{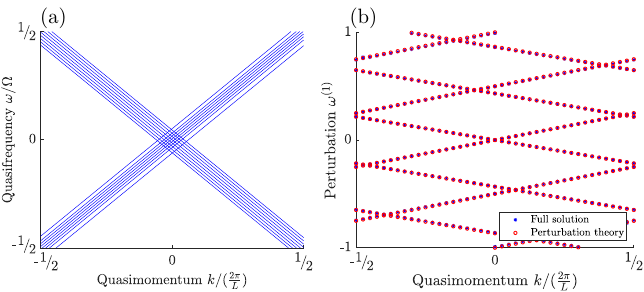}
		\caption{The perturbation theory may be applied to travelling-wave modulations. Here, we consider the luminal case $\alpha = 1$ and $\epsilon = 0.01$, where the full band structure is shown in (a). Again, we observe close agreement between the perturbations $\omega^{(1)}$, computed either through the full solution of \eqref{eq:spatial} or through the perturbation theory \eqref{eq:result}, shown in (b).}\label{fig:travelling}
	\end{figure}

	To exemplify the travelling-wave modulation, we consider the luminal case $v=\frac{\Omega}{g}=1$, which corresponds to $\alpha =1$.	In FIG.~\ref{fig:travelling} we compute the band structure for 
	\begin{equation}
		f(s) = 1+\frac{1}{2}\cos(s).
	\end{equation}	
	In this case, the condition \eqref{eq:assump} holds; consequently, the eigenmodes are exponentially localized in the Fourier domain and may be accurately computed using the perturbation theory. As expected, we observe close agreement between the perturbation theory and the exact solution (computed using the moving coordinate frame).

	\section{Higher-order degeneracies}\label{sec:higher}
	We now derive the perturbation theory close to band intersections of the free dispersion. At these points, marked in FIG. \ref{fig:folding}(d), we have higher degeneracies and a different perturbation theory. The higher-order degeneracies are characterised as intersections between branches of the free dispersion relation separated by frequency $d\Omega$ for some integer $0<d<q$. At order $O(1)$, the solution to \eqref{eq:e0} is now given by
	\begin{equation}
		u_n^{(0)}(x) = \begin{cases} a_{j} v_j(x), \quad &n = qj, \\ b_j w_j(x), & n = qj + d,\\ 
			0,& \text{otherwise},\end{cases}
	\end{equation}
	for the free eigenbasis
	\begin{equation}
		v_j(x) := e^{\iu k_j x}, \quad w_j(x):= e^{-\iu k_j' x},
	\end{equation}
	where $k_j = \frac{\omega^{(0)}+qj\Omega}{c}, \ k_j' = \frac{\omega^{(0)}+(qj+d)\Omega}{c}$ and $\omega^{(0)}(k) = c(k+\frac{2\pi \ell}{L})$, for $\ell\in \Z$ and $j \in \Z$. Projecting \eqref{eq:Oe} onto this basis, we obtain the $2\times 2$ block system 
	\begin{equation}\label{eq:evalHigh}
		-\frac{1}{2}\begin{pmatrix}\mathbf{H} & \mathbf{G} \\ \mathbf{G}^* & \mathbf{H}			
		\end{pmatrix}\begin{pmatrix}\mathbf{O} & \\  & \mathbf{O}'		
		\end{pmatrix}\mathbf{v} = \omega^{(1)}\mathbf{v},
	\end{equation}
	where 
	\begin{equation}
		\mathbf{v} = \begin{pmatrix}\mathbf{O} & \\  & \mathbf{O}'		
		\end{pmatrix}\begin{pmatrix}\mathbf{a} \\ \mathbf{b} \end{pmatrix}, \quad \mathbf{a} = \begin{psmallmatrix} \scalebox{.5}{$\vdots$} \\[-0.1em] a_{-1}\\a_0\\a_1\\[-0.2em] \scalebox{.5}{$\vdots$} \end{psmallmatrix}, \ \mathbf{b} = \begin{psmallmatrix} \scalebox{.5}{$\vdots$} \\[-0.1em] b_{-1}\\b_0\\b_1\\[-0.2em] \scalebox{.5}{$\vdots$} \end{psmallmatrix}.
	\end{equation}
	Here, $\mathbf{H}$ and $\mathbf{O}$ are defined as before, $\mathbf{O}'$ is the diagonal matrix with entries
	\begin{equation}
		O'_j = \omega^{(0)} + (qj+d)\Omega,
	\end{equation}
	and the entries of the off-diagonal block $\mathbf{G}$ are given by non-integer Fourier coefficients of $\eta$:
	\begin{align}
		G_{ij}  &= \ds\dfrac{1}{L}\int_{0}^{L}e^{-\iu \left( \frac{2\omega^{(0)}}{c} +p\left(i+j+\frac{d}{q}\right) \frac{2\pi}{L}\right)x}\hat{\eta}(x,q(j-i)-d) \dx x,\\
		&= \widetilde{\eta}\left(\tfrac{\omega^{(0)}L}{c\pi} +p\left(i+j+\tfrac{d}{q}\right),q(j-i)-d\right),
	\end{align}
	for $i,j\in \Z$. 
	
\begin{figure}[tbh]
		\centering
		\includegraphics[width=\linewidth]{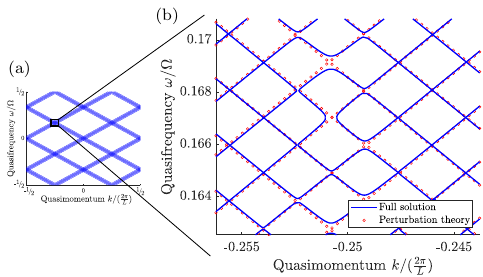}
		\caption{We consider the same numerical experiment as in FIG. \ref{fig:numerics1} and restrict attention to the higher-order degenerate point marked in (a). There are two families of perturbed bands, computed in (b) using the perturbation theory of Section \ref{sec:higher}. The full solution show higher-order characteristics not captured by this first-order perturbation theory: around the intersections, the bands open to local band gaps and momentum gaps.}\label{fig:numerics2}
	\end{figure}

	FIG. \ref{fig:numerics2} shows the perturbed bands in a neighbourhood of a higher-order degeneracy. Although the main characteristics agree, higher-order interactions of the full solution lead to level repulsion and the emergence of local band gaps and momentum gaps. This is not captured by the first-order perturbation theory, which does not account for hybridisation between the perturbed bands.  {In other words, the width of the gaps are of order $O(\epsilon^2)$, and can only be characterised using a higher-order perturbation theory.}

	\section{Irrational band folding}
	 {We next consider the case when $\alpha$ is irrational. We can adopt a rational approximation}
	\begin{equation}
		\alpha = \lim_{i\to \infty} \frac{p_i}{q_i},
	\end{equation}
	for a sequence of integers $p_i,q_i \in \Z$ with $p_i,q_i \xrightarrow{i\to \infty} \infty$. Since $\widetilde{\eta}(m,n) \to 0 $ as $m\to \infty$ or $n\to \infty$, the matrix $\mathbf{H}$ tends to a diagonal matrix as $i\to \infty$. In the limit, although each band of the free dispersion relation is arbitrarily close to other bands, the coupling terms decay as the rational approximation improves. 
	
\begin{figure}[tbh]
		\centering
		\includegraphics[width=\linewidth]{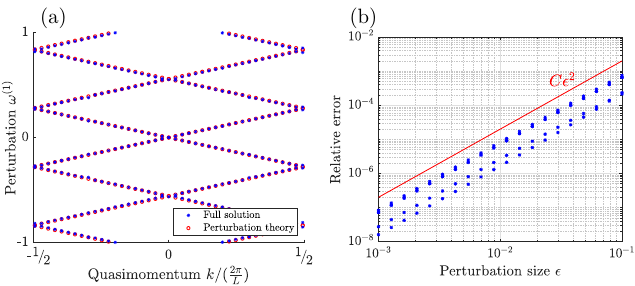}
		\caption{For irrational $\alpha$, the folded bands decouple and may be computed using a finitely degenerate perturbation theory. We consider the same numerical experiment as in FIG.~\ref{fig:numerics1} but using the value $\alpha = \sqrt{2}$. In (a), we show the first perturbation terms computed using \eqref{eq:evalfinite}. In (b), we fix $k=0.5(\frac{2\pi}{L})$ and observe the expected quadratic error as $\epsilon \to 0$.}\label{fig:irrational}
	\end{figure}

	We can compute the band structure using a $2$-fold (finite) degenerate perturbation theory. The solution to \eqref{eq:e0} is now given by
	\begin{equation}
		u_n^{(0)}(x) = \begin{cases} av(x), & n = n_1,\\ bw(x), & n = n_2,\\ 0, & \text{otherwise},\end{cases}
	\end{equation}
	where $v(x) = e^{\iu k_1 x}$ and $w(x) = e^{-\iu k_2 x}$ for $k_i = \frac{\omega^{(0)} + n_i\Omega}{c}$. Moreover, $n_1$ and $n_2$ are integers such that $k_1 = k+\ell_1 g$ and $k_2 = -k+\ell_2 g$ for some integers $\ell_1$ and $\ell_2$. Projecting \eqref{eq:Oe} onto this basis, we obtain the $2\times 2$ (finite) system 
	\begin{equation}\label{eq:evalfinite}
		-\frac{1}{2}\begin{pmatrix}H_0 & G \\ G^* & H_0		
		\end{pmatrix}\begin{pmatrix} \omega^{(0)} + n_1\Omega & \\  & \omega^{(0)} + n_2\Omega		
		\end{pmatrix}\mathbf{v} = \omega^{(1)}\mathbf{v},
	\end{equation}
	where $G$ is given by
	\begin{equation}
		G  = \widetilde{\eta}\left( \tfrac{\omega^{(0)}L}{c\pi} + (n_1+n_2)\alpha,n_1-n_2\right).
	\end{equation}
	We illustrate the irrational case numerically in FIG.~\ref{fig:irrational} using the same setup as in FIG.~\ref{fig:numerics1} but with $\alpha = \sqrt{2}$. Notably, in (b), we observe that the error scales as $O(\epsilon^2)$ which is consistent with the first-order perturbation theory developed herein.
	
	\section{Conclusions}
	We have developed a perturbation theory for band structures of spacetime-periodic structures. Due to the infinite-degenerate nature of the free problem, the effective equations are given by an eigenvalue problem of an infinite matrix. Computing these eigenvalues, we find that the infinitely-degenerate bands may split into a family of non-degenerate bands.  {We have validated our method by considering two simplified settings where full, numerically computed solutions are readily available, and we observe close agreement.}
	
	The formal asymptotic expansions of this work, while demonstrating excellent agreement with numerical methods, open a rich field of mathematical questions. Proving the validity of the formal expansions, and characterising the spectrum of the limiting problem, remain open problems. Specifically, the case when \eqref{eq:assump} no longer holds will be of significant interest. In this case, we may have strong coupling between modes of different frequencies, and the solutions are no longer exponentially localized in the Fourier domain. {Moreover, computing the higher-order correction terms will be of fundamental importance for describing the emergence of momentum gaps and non-reciprocal band gaps.}
	
	We conclude by remarking that the methods are general with respect to the form of the equation \eqref{eq:wave}. Although phrased herein for a generic scalar wave equation, the methods are applicable to specific one-dimensional models for photonics and phononics, where the main difference would be an altered form of the eigenvalue problem \eqref{eq:eval}.
		
	\bibliography{references}
	
\end{document}